\documentclass[12pt]{iopart}
\usepackage{graphicx}

\begin{document} 

\def\Journal#1#2#3#4{{#1} {\bf #2}, #3 (#4)}

\def\NCA{Nuovo Cimento}
\def\NIM{Nucl. Instr. Meth.}
\def\NIMA{{Nucl. Instr. Meth.} A}
\def\NPB{{Nucl. Phys.} B}
\def\NPA{{Nucl. Phys.} A}
\def\PLB{{Phys. Lett.}  B}
\def\PRL{Phys. Rev. Lett.}
\def\PRC{{Phys. Rev.} C}
\def\PRD{{Phys. Rev.} D}
\def\ZPC{{Z. Phys.} C}
\def\JPG{{J. Phys.} G}
\def\CPC{Comput. Phys. Commun.}
\def\EPJ{{Eur. Phys. J.} C}

\title[Open Charm Yields in 200 GeV p+p and d+Au Collisions at RHIC]{Open Charm Yields in 200 GeV p+p and d+Au Collisions at RHIC}

\author{Lijuan Ruan\footnote[1]{rlj@mail.ustc.edu.cn} (for the STAR Collaboration\footnote[2]{For the full author list and acknowledgements see Appendix ``Collaborations'' in this volume.})}
\address{Dept. of Modern Physics, University of Science and Technology of China, Hefei, Anhui, China, 230026;
Brookhaven National Laboratory, Upton, NY, 11973, USA}

\begin{abstract}
Open charm spectra at mid-rapidity from direct reconstruction of
$D^{0}\rightarrow K\pi$ and indirect electron/positron measurements via
charm semileptonic decay in p+p and d+Au collisions at $\sqrt{s_{_{\rm NN}}}$ = 200 GeV
are reported. The combined spectra cover open charm transverse
momentum range $0.1{}^{<}_{\sim}p_{T}{}^{<}_{\sim}6$ GeV/$c$. The electron 
spectra show approximate binary scaling between p+p and d+Au collisions.
From these two independent
analyses, the differential cross section per
nucleon-nucleon interaction at mid-rapidity for open charm production
at RHIC is $d\sigma^{NN}_{c\bar{c}}/dy$=0.31$\pm$0.04(stat.)$\pm$0.08(syst.) mb.
The next-to-leading-order pQCD
calculation underpredicts the charm yields.
\end{abstract}

The production and spectra of hadrons with heavy flavor are sensitive to 
initial conditions and the later stage dynamical evolution in high energy
 nuclear collisions, and may be
less affected by the non-perturbative complication in theoretical
calculations \cite{charm1}. Charm production has been proposed as a
sensitive measurement of parton distribution function in nucleon and
the nuclear shadowing effect by systematically studying p+p, and
p+A collisions \cite{lin96}. The relatively reduced
energy loss of heavy quark traversing a Quark-Gluon Plasma will help us distinguish the medium in which the jet loses its
energy~\cite{dokshitzer01}. A possible
 enhancement of charmonia ($J/\Psi$) production can be present at RHIC energies
 \cite{jpsi} due to the coalescence of the copiously produced charm quarks.

Direct and indirect charm production has been measured in hadron-hadron 
collisions \cite{isrc,fermic,phenixe}. However, no experimental results are available at $\sqrt{s_{_{\rm NN}}}$=200GeV.
Theoretical predictions for this energy region differ
 significantly \cite{vogt02}. Therefore precise experimental measurement
 of the baseline charm cross sections at RHIC is necessary. In this
 paper, we report the recent STAR results on the
absolute open charm cross section measurements from direct charmed
hadron $D^0$ reconstruction in d+Au
collisions and electrons from charm semileptonic decay in both p+p and
d+Au collisions at 200 GeV.
\section{Analysis methods and results}
\begin{figure} [htb]
\begin{center}
\includegraphics[width=0.9\textwidth]{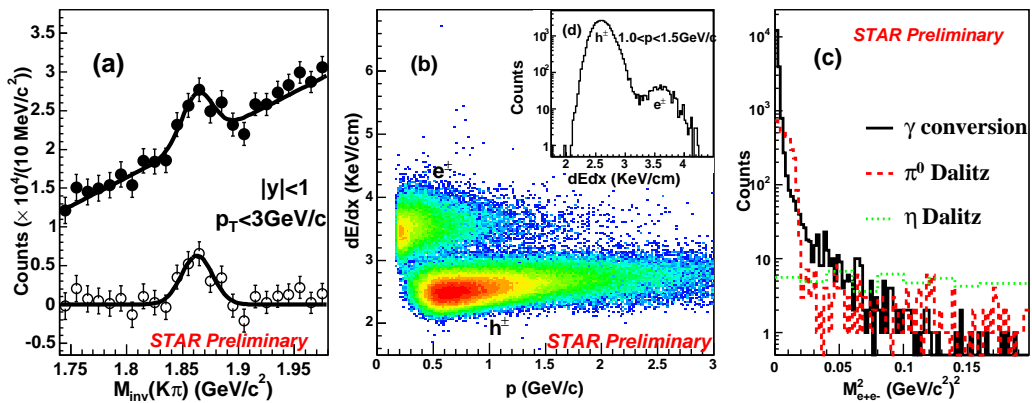} 
\caption{(a) Invariant mass
 distributions of oppositely charged $K\pi$ pairs from d+Au
 collisions. (b) $dE/dx$ in TPC versus $p$ with the TOFr velocity cut $|1/\beta- 1|\le0.03$. The insert shows $dE/dx$ distribution for 1 $\le p \le$ 1.5 GeV/c. (c) The $e^{+}e^{-}$ pair invariant mass square distribution from different sources in PYTHIA \cite{pythia1} simulation. }
\end{center}
\label{fig1pid}
\end{figure}

  At STAR, the hadronic channel of $D^{0} \rightarrow K^{-}\pi^{+} (+c.c.)$ with branching ratio of 3.83\% was measure by the Time Projection
  Chamber (TPC) \cite{startpctof,stardau} in d+Au collisions. The $D^{0}$ signal was reconstructed using the event-mixing technique which is the same as the one used to identify and measure the $K^{0*}$ resonance \cite{kstarpaper}. 
After the mixed-event background subtraction, the $D^0$ signal in $p_T<3$
GeV/c and $|y|<1$ from 15.7 million d+Au minimum bias events is shown as filled symbols in
Fig. 1(a).
 A linear function was used to reproduce the residual background. After the 
two-step background subtraction, a gaussian function was used to obtain 
the $D^0$ signal shown as open symbols.

In the 2002-2003 run, a prototype multi-gap resistive plate chamber time-of-flight system (TOFr) \cite{startpctof} was
installed with the coverage $-1\!<\!\eta\!<\!0$ in pseudorapidity.
In addition to its capability of hadron
identification \cite{startof1}, electrons could be identified at low
momentum ($p_{T}\le3$ GeV/c) by the combination of velocity ($\beta$) from TOFr \cite{startpctof}
and the
 particle ionization energy loss ($dE/dx$) from TPC \cite{startpctof}. 
Fig. 1(b) shows that the electrons are clearly identified as
a separate band in the $dE/dx$ versus momentum ($p$) with a selection on $\beta$ at $|1/\beta-1|\le0.03$. At
higher $p_{T}$ (2--4 GeV/c), negative electrons were also identified directly by TPC
since hadrons have lower $dE/dx$ due to the relativisitic rise of
electron $dE/dx$. Detector acceptance and efficiency corrections were determined from
the embedding data \cite{startof1}. Total inclusive electron spectra from p+p
and d+Au collisions are shown as symbols in Fig. 2.

\begin{figure}[htb] \centerline{\includegraphics[width=0.475\textwidth]
{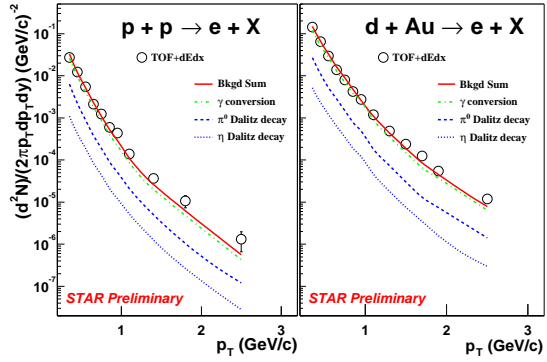}} 
\vspace{-0.35cm}

\caption[]{Electron distributions from p+p (left) and d+Au
collisions (right), respectively. The symbols are
electrons identified via TOFr combined with TPC.}

\label{fig3einc} \end{figure}

$\gamma$ conversions and $\pi^0$ Dalitz decays are the dominant photonic sources
for electron background. Such background was mainly constrained at
small pair invariant mass and small opening angle.  To measure the background,
pair invariant mass and opening angle distributions were constructed
by first selecting an electron from TOFr and then looping through
oppositely charged tracks reconstructed in the TPC \cite{johnson}. Fig. 1(c) shows an example of the pair invariant mass.  Simulations using both HIJING \cite{hijing} and PYTHIA \cite{pythia1}, with
GEANT to describe the detector, showed $\sim$60\% effeciency of such
background reconstruction for $p_{T}$ $\ge 1$GeV/c, with negligible $p_{T}$ dependence. 
This value was used to correct for the photonic
electron spectra in data. In this method, more than 95\% background
has been measured, shown as thick solid lines in
Fig. 2. Other contributions ($<5\%$) from decays of 
$\eta, \omega, \rho, \phi$ and $K$ were determined from simulations. The overall uncertainty of the
background is $\sim$20\%.

The non-photonic electron spectra, obtained by subtracting the
previously described photonic background from the inclusive spectra,
are shown as
symbols in Fig. 3 (left) for both p+p (triangles) and d+Au (circles)
collisions. 
The $D^0$
invariant yields $d^2N/2\pi p_Tdp_Tdy$ are calculated and also shown in
Fig. 3 (left) as squared symbols as a function of $p_T$. The
mid-rapidity yield $dN/dy$ was extracted with an exponential fit to the
invariant yield as a function of transverse mass $m_T$, which is $0.028 \pm0.004 \pm0.007$.  
Several model 
studies \cite{charm1,pythia1} showed that semi-leptonic decays from open
charm are the dominant non-photonic electrons at 1 $\le$ $p_T$
$\le 4$ GeV/c.
We performed a
fit with the combined results of $D^0$ and electron distributions in d+Au collisions, assuming that the $D^0$ spectrum follows the power law up to 6 GeV/c. From the fit, we got the yield of $D^{0}$ at mid-rapidity $dN/dy=0.030\pm0.004\pm0.008$ and $d\sigma_{c\bar{c}}^{NN}$/dy$=0.31\pm0.04\pm0.08$ mb. 
Thus the total charm-pair cross section $\sigma_{c\bar{c}}^{NN}=1.44\pm0.20\pm0.44$ mb, where a factor of
$4.7\pm0.7$, estimated by the model simulations \cite{vogt02,pythia1}, was used to convert the $d\sigma/dy$ at mid-rapidity to total cross section. The systematic error is
 dominated by the background subtraction, $p_T$ coverage for each measurement and overall normalization. The nuclear
 modification factor was obtained by taking the ratio of the electron
 spectra in Fig. 3 (left) scaled with the underlying
 nucleon-nucleon binary collisions. It was measured to be
 $1.2\pm0.2\pm0.3$ and is consistent with binary scaling within the
 errors.

\begin{figure}[h] \centerline{\hspace{3.0cm}
\includegraphics[width=0.4\textwidth]
{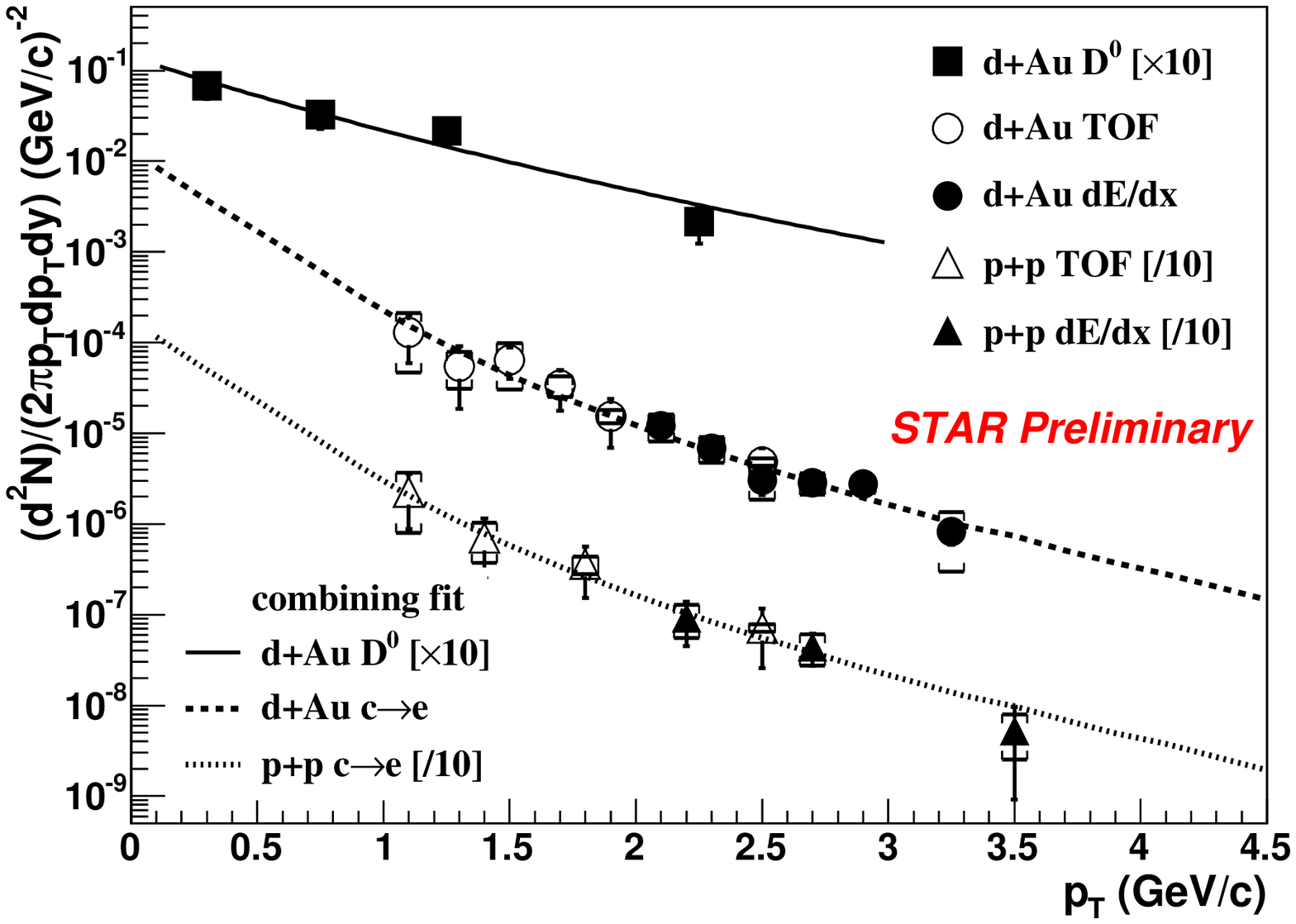}\includegraphics
[width=0.45\textwidth, height=5.0cm]{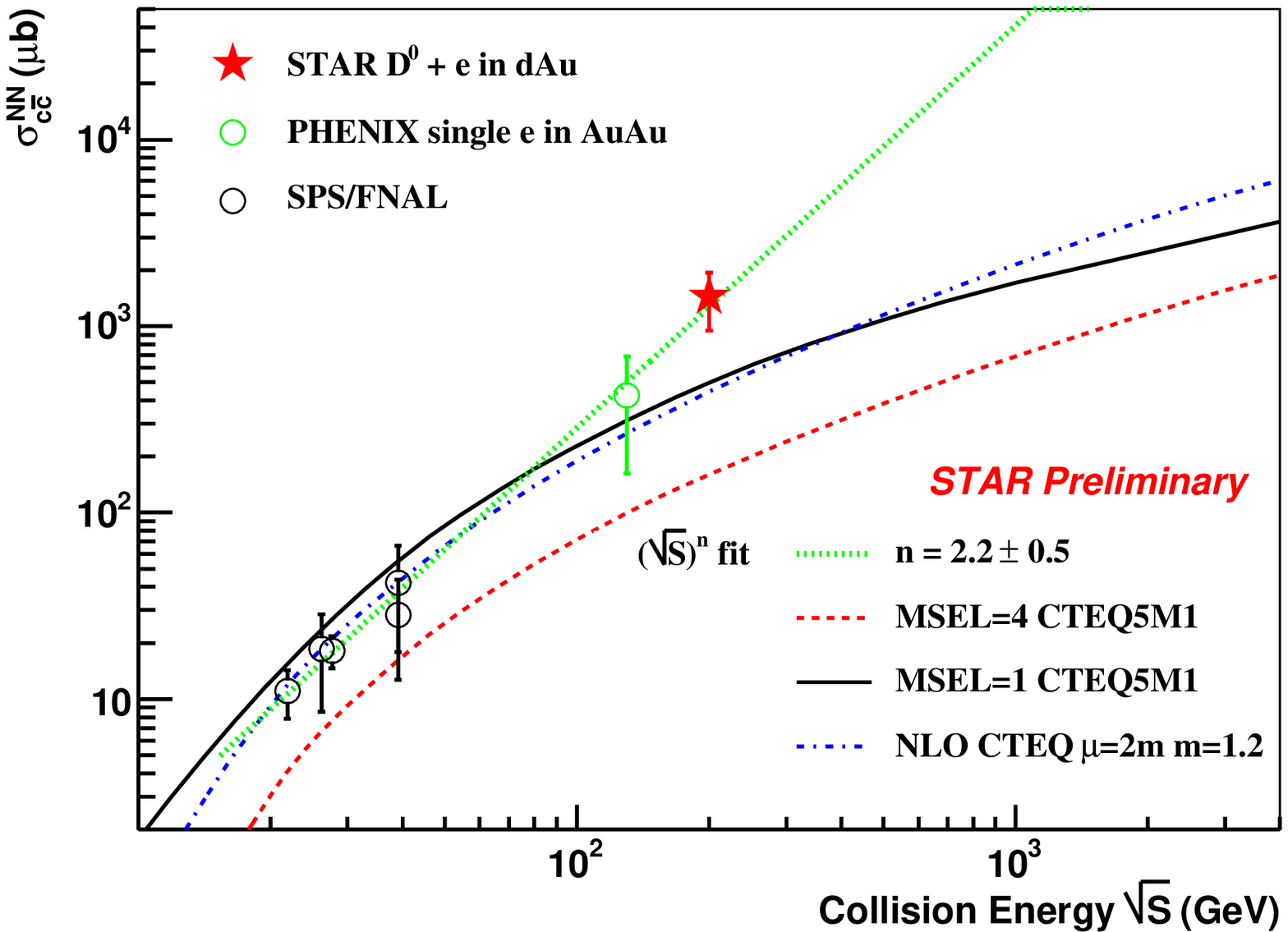}}
\vspace{-0.35cm}

\caption[]{ (left) Reconstructed $D^0$ $p_T$ distributions
from d+Au collisions. Non-photonic electron $p_T$ distributions 
from p+p and d+Au collisions (open and filled symbols are
electrons identified via TOFr combined with TPC and TPC only). Solid and dashed lines are the combined fit results for the $D^0$ and electron spectra in d+Au collisions. Dotted line is scaled down by the number of binary collisions ($N_{bin}$=7.5) in d+Au collisions. (right) Total charm-pair cross sections 
versus the collision energy (in $\sqrt{s_{_{\rm NN}}}$). The result of our measurement is shown as star.  
}
\label{fig4de}
\end{figure}

The beam energy dependence of the cross section
 is shown in Fig. 3 (right). The solid (MSEL=1) and dashed (MSEL=4) lines are PYTHIA calcualtions with and without high order processes such as flavor excitation $etc.$, respectively. The dot-dashed line
is the NLO calculation from \cite{vogt02}, which underpredicts our result.
 Dotted line is the power-law
 fit of $\sigma_{c\bar{c}}^{NN}\propto(\sqrt{s})^n$ to the data points. 
For the total charm production, the power $n\sim2.2\pm0.5$, while $n \sim 0.5 (0.3)$ has been observed for pion (charged multiplicity) productions at lower energy \cite{jpsi,na49pi}.

\section{Summary}

The charm cross section, $d\sigma^{NN}_{c\bar{c}}/dy$=0.31$\pm$0.04(stat.)$\pm$0.08(syst.) mb from 200 GeV d+Au collisions has been
measured at RHIC. The independent measurements
of the reconstructed $D^0$ and single electrons from charm semileptonic decay are consistent. The NLO calculation underpredicts the total
cross section at this energy.


\vspace{0.35cm}

\section*{References}

\end{document}